\documentclass[11pt,a4paper]{article}
\usepackage{amsmath,amssymb}
\usepackage{epsfig,graphicx}

\input babarsym

\usepackage{relsize}
\def\babar{\mbox{\slshape B\kern-0.1em{\smaller A}\kern-0.1em
    B\kern-0.1em{\smaller A\kern-0.2em R}}}

\def\en         {\ensuremath{e^-}\xspace}   
\def\ep         {\ensuremath{e^+}\xspace}

\def\mup        {\ensuremath{\mu^+}\xspace}
\def\mun        {\ensuremath{\mu^-}\xspace} 
\def\mumu       {\ensuremath{\mu^+\mu^-}\xspace}

\def\taup       {\ensuremath{\tau^+}\xspace}
\def\taum       {\ensuremath{\tau^-}\xspace}

\def\ellell     {\ensuremath{\ell^+ \ell^-}\xspace}

\def\nub        {\ensuremath{\overline{\nu}}\xspace}
\def\nunub      {\ensuremath{\nu{\overline{\nu}}}\xspace}
\def\nub        {\ensuremath{\overline{\nu}}\xspace}
\def\nunub      {\ensuremath{\nu{\overline{\nu}}}\xspace}
\def\nue        {\ensuremath{\nu_e}\xspace}
\def\nueb       {\ensuremath{\nub_e}\xspace}

\def\num        {\ensuremath{\nu_\mu}\xspace}
\def\numb       {\ensuremath{\nub_\mu}\xspace}

\def\nut        {\ensuremath{\nu_\tau}\xspace}
\def\nutb       {\ensuremath{\nub_\tau}\xspace}

\def\nul        {\ensuremath{\nu_\ell}\xspace}
\def\nulb       {\ensuremath{\nub_\ell}\xspace}

\def\g     {\ensuremath{\gamma}\xspace}
\def\gaga  {\ensuremath{\gamma\gamma}\xspace}  
\def\ggstar{\ensuremath{\gamma\gamma^*}\xspace}

\def\t     {\ensuremath{t}\xspace}
\def\tbar  {\ensuremath{\overline t}\xspace}
\def\tbar  {\ensuremath{\overline t}\xspace}

\def \tb     {\ensuremath{\tbar}\xspace}
\def \ttb    {\ensuremath{\t {\kern -0.16em \tb}}\xspace}

\def\piz   {\ensuremath{\pi^0}\xspace}

\def\ppz   {\ensuremath{\pi^0\pi^0}\xspace}
\def\pip   {\ensuremath{\pi^+}\xspace}
\def\pim   {\ensuremath{\pi^-}\xspace}
\def\pipi  {\ensuremath{\pi^+\pi^-}\xspace}

\def\pimp  {\ensuremath{\pi^\mp}\xspace}

\def\kaon  {\ensuremath{K}\xspace}
\def\Kbar  {\kern 0.2em\overline{\kern -0.2em K}{}\xspace}

\def\Kz    {\ensuremath{K^0}\xspace}
\def\Kzb   {\ensuremath{\Kbar^0}\xspace}
\def\KzKzb {\ensuremath{\Kz \kern -0.16em \Kzb}\xspace}
\def\Kp    {\ensuremath{K^+}\xspace}
\def\Km    {\ensuremath{K^-}\xspace}
\def\Kpm   {\ensuremath{K^\pm}\xspace}

\def\KpKm  {\ensuremath{\Kp \kern -0.16em \Km}\xspace}
\def\KS    {\ensuremath{K^0_{\scriptscriptstyle S}}\xspace} 
\def\KL    {\ensuremath{K^0_{\scriptscriptstyle L}}\xspace} 
\def\Kstarz  {\ensuremath{K^{*0}}\xspace}

\def\Kstar   {\ensuremath{K^*}\xspace}

\def\Kstarp  {\ensuremath{K^{*+}}\xspace}

\def\Dbar    {\kern 0.2em\overline{\kern -0.2em D}{}\xspace}

\def\Dz      {\ensuremath{D^0}\xspace}
\def\Dzb     {\ensuremath{\Dbar^0}\xspace}
\def\DzDzb   {\ensuremath{\Dz {\kern -0.16em \Dzb}}\xspace}
\def\Dp      {\ensuremath{D^+}\xspace}
\def\Dm      {\ensuremath{D^-}\xspace}

\def\DpDm    {\ensuremath{\Dp {\kern -0.16em \Dm}}\xspace}
\def\Dstar   {\ensuremath{D^*}\xspace}

\def\Dstarz  {\ensuremath{D^{*0}}\xspace}

\def\Dstarp  {\ensuremath{D^{*+}}\xspace}
\def\Dstarm  {\ensuremath{D^{*-}}\xspace}

\def\B       {\ensuremath{B}\xspace}
\def\Bpr     {\ensuremath{B^{*}}\xspace}
\def\Bs      {\ensuremath{B_{s}}\xspace}
\def\Bss     {\ensuremath{B_{s}^{*}}\xspace}
\def\Bsss    {\ensuremath{B_{s}^{(*)}}\xspace}

\def\Bbar    {\kern 0.18em\overline{\kern -0.18em B}{}\xspace}
\def\Bprb    {\kern 0.18em\overline{\kern -0.18em B^{*}}{}\xspace}
\def\BBpr    {\ensuremath{\B {\kern -0.16em \Bpr}}\xspace}
\def\BBprb   {\ensuremath{\B {\kern -0.16em \Bprb}}\xspace}
\def\BprBprb {\ensuremath{\Bpr {\kern -0.16em \Bprb}}\xspace}
\def\Bsbar   {\kern 0.18em\overline{\kern -0.18em B_{s}}{}\xspace}
\def\Bssbar  {\kern 0.18em\overline{\kern -0.18em B_{s}^{*}}{}\xspace}
\def\Bsssbar {\kern 0.18em\overline{\kern -0.18em B_{s}^{(*)}}{}\xspace}
\def\Bb      {\ensuremath{\Bbar}\xspace}

\def\BB      {\ensuremath{\B {\kern -0.16em \Bb}}\xspace}
\def\BBp     {\ensuremath{\B {\kern -0.16em \Bb}\pi}\xspace}
\def\BBpp    {\ensuremath{\B {\kern -0.16em \Bb}\pi\pi}\xspace}
\def\BBprp   {\ensuremath{\B {\kern -0.16em \Bprb}\pi}\xspace}
\def\BsBsb   {\ensuremath{\Bs {\kern -0.16em \Bsbar}}\xspace}
\def\BsBssb  {\ensuremath{\Bs {\kern -0.16em \Bssbar}}\xspace}
\def\BssbBs  {\ensuremath{\Bssbar {\kern -0.16em \Bs}}\xspace}
\def\BssBsb  {\ensuremath{\Bss {\kern -0.16em \Bsbar}}\xspace}
\def\BssBssb {\ensuremath{\Bss {\kern -0.16em \Bssbar}}\xspace}
\def\Bz      {\ensuremath{B^0}\xspace}
\def\Bsz     {\ensuremath{B_{s}^0}\xspace}
\def\Bzb     {\ensuremath{\Bbar^0}\xspace}
\def\Bszb    {\ensuremath{\Bsbar^0}\xspace}
\def\BzBzb   {\ensuremath{\Bz {\kern -0.16em \Bzb}}\xspace}
\def\BszBszb {\ensuremath{\Bsz {\kern -0.16em \Bszb}}\xspace}
\def\BsssBsssb {\ensuremath{\Bsss {\kern -0.16em \Bsssbar}}\xspace}

\def\Bu      {\ensuremath{B^+}\xspace}
\def\Bub     {\ensuremath{B^-}\xspace}

\def\Bpm     {\ensuremath{B^\pm}\xspace}

\def\BpBm    {\ensuremath{\Bu {\kern -0.16em \Bub}}\xspace}
\def\Bs      {\ensuremath{B_s}\xspace}

\def\BorBbar    {\kern 0.18em\optbar{\kern -0.18em B}{}\xspace}
\def\DorDbar    {\kern 0.18em\optbar{\kern -0.18em D}{}\xspace}
\def\KorKbar    {\kern 0.18em\optbar{\kern -0.18em K}{}\xspace}

\def\jpsi     {\ensuremath{{J\mskip -3mu/\mskip -2mu\psi\mskip 2mu}}\xspace}

\mathchardef\Upsilon="7107
\def\Y#1S{\ensuremath{\Upsilon{(#1S)}}\xspace}

\def\FourS {\Y4S}

\mathchardef\Deltares="7101
\mathchardef\Xi="7104
\mathchardef\Lambda="7103
\mathchardef\Sigma="7106
\mathchardef\Omega="710A

\def\Deltabar{\kern 0.25em\overline{\kern -0.25em \Deltares}{}\xspace}
\def\Lbar{\kern 0.2em\overline{\kern -0.2em\Lambda\kern 0.05em}\kern-0.05em{}\xspace}
\def\Sigbar{\kern 0.2em\overline{\kern -0.2em \Sigma}{}\xspace}
\def\Xibar{\kern 0.2em\overline{\kern -0.2em \Xi}{}\xspace}
\def\Obar{\kern 0.2em\overline{\kern -0.2em \Omega}{}\xspace}
\def\Nbar{\kern 0.2em\overline{\kern -0.2em N}{}\xspace}
\def\Xb{\kern 0.2em\overline{\kern -0.2em X}{}\xspace}

\def\X {\ensuremath{X}\xspace}

\def\BR         {{\ensuremath{\cal B}\xspace}}

\def\bpsikl     {\ensuremath{\Bz \to \jpsi \KL}\xspace}

\def\Bzbtox     {\ensuremath{\Bzb \to \X}\xspace}

\def\mus  {\ensuremath{\rm \,\mus}\xspace}

\def\mus        {\ensuremath{\,\mu{\rm s}}\xspace}    

\def\degrees{\ensuremath{^{\circ}}\xspace}

\def\to                 {\ensuremath{\rightarrow}\xspace}

\def\pep2{PEP-II}
\def\BF{$B$ Factory}

\def\gsim{{~\raise.15em\hbox{$>$}\kern-.85em
          \lower.35em\hbox{$\sim$}~}\xspace}
\def\lsim{{~\raise.15em\hbox{$<$}\kern-.85em
          \lower.35em\hbox{$\sim$}~}\xspace}

\def\epstag  {\ensuremath{\varepsilon_{\rm tag}}\xspace}

\def\jetset74   {\mbox{\tt Jetset \hspace{-0.5em}7.\hspace{-0.2em}4}\xspace}

\usepackage{relsize}

\begin{document}

\title{\bf Hot Topics From \babar\ Experiment}

\author{Romulus Godang\footnote{{\bf e-mail}: godang@usouthal.edu or godang@slac.stanford.edu}\\ 
On Behalf of the \babar\ Collaboration\\
\\
\small{\em Department of Physics} \\
\small{\em University of South Alabama} \\
\small{\em ILB 115, 307 University Blvd., N.} \\
\small{\em Mobile, Alabama 36688}
}
\date{}

\maketitle

\begin{abstract}
We search for a new light non-Standard Model $CP$-odd Higgs boson, $A^0$, decaying to
$\tau^+\tau^-$, $\mu^+\mu^-$, and invisible in radiative decays of the $\Upsilon(2S)$ 
and $\Upsilon(3S)$. We search for the light scalar particle produced in single-photon decays 
of the $\Upsilon(3S)$ resonance through the process $\Upsilon(3S) \to \gamma A^0, A^0 \to$ invisible.
We also search for evidence of an undetectable $\Upsilon(1S)$ decay
recoiling against the dipion system. The data samples contain $99 \times 10^6~\Upsilon(2S)$ 
and $122 \times 10^6~\Upsilon(3S)$ decays collected by the \babar\ detector at the SLAC PEP-II 
$B$ factory.\\ 
\begin{center}
Contributed to the Proceedings of $16^{th}$ International Seminar on High Energy Physics\\ 
QUARKS-2010, Kolomna, Russia, 6-12 June, 2010.\\
\end{center}

\end{abstract}

\hspace{-0.8cm} SLAC-PUB-14254, UMISS-HEP-2010-03

\section{Introduction}

The search for the origin of mass of fundamental particle is a cornerstone of the Standard Model (SM). 
The Higgs mechanism is one of the great interests in the particle physics. In recent years many 
motivations for searching a Higgs boson. Direct searches at the CERN $e^+-e^-$ Collider (LEP Experiment) 
via the process $e^+-e^- \to Z H$ for the Standard Model Higgs boson, a $CP$-even scalar, has excluded 
a lower bound on its mass below 114.4 GeV~\cite{LEP_2003}. The search for the SM Higgs particle is 
continued at the Tevatron (The CDF and D0 Collaborations) for a Higgs boson decaying to $W^+W^-$ and
resulting limits on Higgs boson production exclude a SM Higgs boson in the mass range 162-166 GeV 
at the 95\% C.L~\cite{CDF_D0_2010}.  The search is currently continued at Large Hadron Collider (LHC) 
with the relative increase in charged-particle multiplicity from $\sqrt s = 0.9$ to $7$ 
TeV~\cite{CMS_2010, ATLAS_2010}.

In recent years a number of theoretical models predicted for the existence of a light $CP$-odd Higgs 
boson ($A^0$) related to the Next-to-Minimal Supersymmetric Model (NMSSM). Direct searches constrain 
the mass of $m_{A^0} < 2m_b$, where $m_b$ is the $b$ quark and the decay of $A^0 \to bb$ 
is forbidden~\cite{Gunion_2006}. Of particular interest is to search the lightest $CP$-odd 
Higgs boson in Upsilon decays such as $\Upsilon(1S)$ and $\Upsilon(3S)$. In this paper we \
present a search of the $CP$-odd Higgs boson decaying to $\tau^+\tau^-$, $\mu^+\mu^-$, 
and invisible. The large data sets available at \babar\ experiment allow us to place 
stringent constraints on such theoretical models.

\section{Data Sample}

The data used in this paper were collected with the \babar\ detector at the PEP-II asymmetric-energy
$e^+-e^-$ storage ring at the SLAC National Accelerator Laboratory , operating at $\Upsilon(2S)$ and 
$\Upsilon(3S)$. For searching for $A^0$ decays to either $\tau^+\tau^-$ or $\mu^+\mu^-$, we use 
a data sample of $122 \times 10^6~\Upsilon(3S)$ events that corresponds to an integrated luminosity 
of 28 $fb^{-1}$. We also used a data sample of 79 $fb^{-1}$ accumulated on the $\Upsilon(4S)$ 
resonance (ON-resonance) and 8 $fb^{-1}$ 49 MeV below the $\Upsilon(4S)$ resonance (OFF-resonance) 
for studying the background. 

For searching an undetectable $\Upsilon(1S)$ decay recoiling against the dipion system we used 
the data were taken using an upgraded muon system, instrumented with both resistive plate chambers 
and the limited streamer tubes between between steel absorbers. The data triggers was modified to 
substantially increase the pion trigger efficiency. The data sample of $96.5 \times 10^{-6}$ $\Upsilon(3S)$ 
were used. The \babar\ detector is described in detail elsewhere~\cite{Babar_nim2002}.

\section{Analysis Method}

\subsection{Search for \boldmath{$A^0$} in the decays of \boldmath {$\Upsilon(3S) \to \gamma A^0$}, 
\boldmath {$A^0 \to \tau^+\tau^-$}}

We search the light $CP$-odd Higgs boson, $A^0$, via the decays $\Upsilon(3S)\to \gamma \tau^+\tau^-$ for 
a wider mass range, $4.03 < m_{\tau^+\tau^-} < 10.10$ GeV/$c^2$~\cite{BABAR_tautau}, over the mass ranges by 
the CLEO Collaboration~\cite{CLEO_tautau} and the D0 Collaboration~\cite{D0_tautau}, respectively. 
We exclude the mass region of $9.52 < m_{\tau^+\tau^-} < 9.61$ GeV/$c^2$ because of the irreducible background 
photons produced in the decays of $\Upsilon(3S) \to \gamma \chi_{bJ}(2P), \chi_{bJ}(2P) \to \gamma \Upsilon(1S)$,
where $J=0, 1, 2$. We scan for peaks in the distribution of the photon energy, $E_{\gamma}$, corresponding 
to peaks in the $\tau \tau$ invariant mass as given by
\begin{eqnarray}
m^2_{\tau^+\tau^-} = m^2_{\Upsilon(3S)} - 2 m_{\Upsilon(3S)}E_{\gamma},
\end{eqnarray}
where $m_{\Upsilon(3S)}$ is the $\Upsilon(3S)$ mass (10.355 GeV/$c^2$) and the $E_{\gamma}$ 
is measured in the $\Upsilon(3S)$ center-of-mass frame.  

We select events in which both $\tau$-leptons decay leptonically to either 
$\tau^+ \to e^+ \nu_{e}\bar{\nu}_{\tau}$ or $\tau^+ \to \mu^+ \nu_{\mu} \bar{\nu}_{\tau}$. 
The events are then required exactly two charged tracks to reduce the background and to contain 
at least one photon with energy $ >$ 100 MeV in the electromagnetic calorimeter. 
In addition both charged tracks are required to be identified as leptons either electron or muon.
The residual background is mostly due to radiative decays of $e^+e^- \to \gamma \tau^+\tau^-$ 
and higher order of QED processes, including two-photon reactions such as 
$e^+e^- \to e^+e^- e^+e^- $ and $e^+e^- \to e^+e^- \mu^+\mu^-$. However this residual background
has smaller contributions compared to other $\Upsilon(3S)$ decays and the continuum background
from the non-resonance $\Upsilon(3S)$ decays of $e^+e^- \to \gamma^* \to q\bar{q}$, where $q = u, d, s, c$. 

\begin{figure}[!htb]
\begin{center}
\vspace*{-0.1cm}
\includegraphics[height=11.0cm]{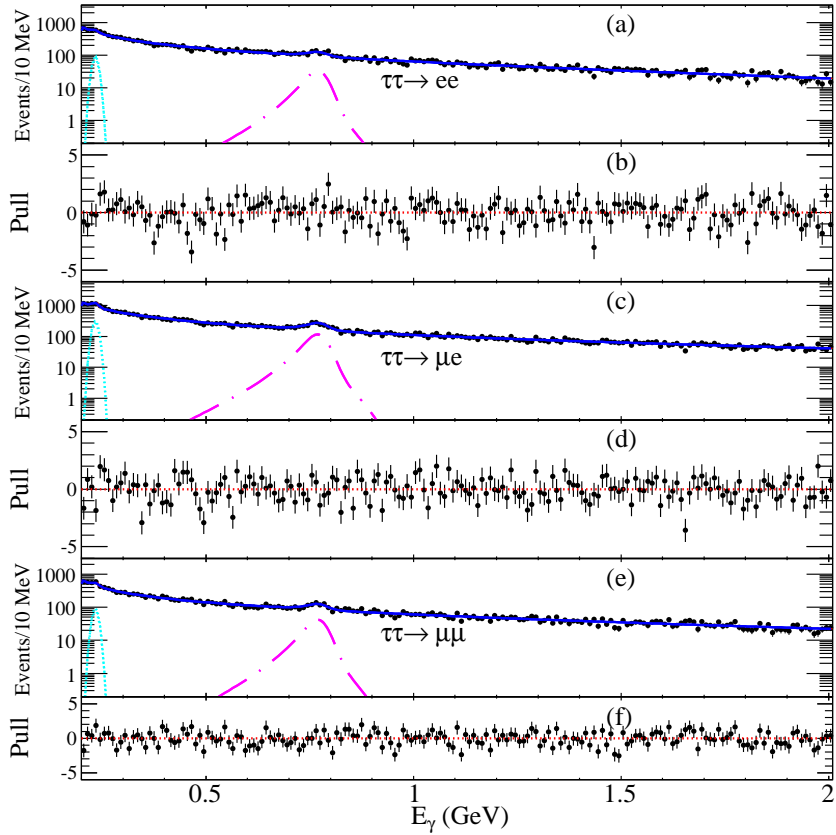}
\vspace*{-0.1cm}
\caption{The photon energy distributions for the different $\tau \tau$-decay modes. The data is shown 
in filled circles, the dotted lines represent the contributions from $\Upsilon(3S) \to \gamma \chi_{bJ}(2P), 
\chi_{bJ}(2P) \to \gamma \Upsilon(2S)$, dotted-dashed lines show the contributions from 
$\Upsilon(3S) \to \gamma \chi_{bJ}(2P), \chi_{bJ}(2P) \to \gamma \Upsilon(1S)$, and the solid lines show
the total background functions. The different between the background function and the data divided by 
the uncertainty in the data for each $\tau \tau$-decay modes is shown in b, c, and d.}  
\label{fig:Egamma_tautau}
\end{center}
\end{figure}

The events $\Upsilon(3S) \to \gamma \chi_{bJ}(2P), \chi_{bJ}(2P) \to \gamma \Upsilon(nS)$, and 
$\Upsilon(nS)$ decays to $\tau^+\tau^-$, where $J = 0, 1, 2$ and $n = 1, 2$ are expected to peak in the 
photon energy distribution when the photon comes from the decays of $\chi_{bJ}(2P) \to \gamma \Upsilon(nS)$
is misidentified as the radiative photon from the $\Upsilon(3S)$ decays. 
We use a Crystal Ball function (CB)~\cite{CB_function} to describes each of the peaks that comes from 
to the decays of $\chi_{bJ}(2P) \to \gamma \Upsilon(1S)$ in the photon spectrum. The mean values for
the $\chi_{b0}(2P)$ and the $\chi_{b1}(2P)$ CB function are fixed to the PDG values~\cite{PDG_2008},
and the width values are fixed to the Monte Carlo (MC) resolution, but the mean and the width of 
$\chi_{b2}(2P)$ are free parameters. The search for the signal $\Upsilon(3S) \to \gamma A^0, 
A^0 \to \tau^+\tau^-$ is performed by scanning for peaks in the photon energy distributions. 
Figure~\ref{fig:Egamma_tautau} shows the fits to the photon energy, $E_{\gamma}$, distributions 
in the different $\tau^+ \tau^-$ decay modes fitted in the region $0.2 < E_{gamma} < 2.0$ GeV.

No evidence for a signature of the light Higgs boson decays to $\tau$ pairs is observed. 
We measure the branching fraction of ${\cal{B}}(\Upsilon(3S) \to \gamma A^0) \times {\cal{B}}
(A^0 \to \tau^+ \tau^-) < (1.5-16) \times 10^{-5}$ \@ 90\% confidence level (C.L.). 
We also set a 90\% C.L. upper limit on the $\tau^+\tau^-$ decays of the $\eta_b$ 
at ${\cal{B}}(\eta_b \to \tau^+ \tau^-)$.  

\subsection{Search for \boldmath{$A^0$} in the decays of 
\boldmath{$\Upsilon(3S) \to \gamma A^0$}, \boldmath{$A^0 \to \mu^+\mu^-$}}

We search for a resonance in the $\mu^+ \mu^-$ invariant mass distribution for the fully reconstructed 
final state $\Upsilon(2S,3S) \to \gamma A^0, A^0 \to \mu^+\mu^-$. We assume the $A^0$ resonance is a scalar
(or pseudoscalar) particle and the width of the resonance is negligibly small compared with the experimental
resolution where the mass of light Higgs boson, $m_{A^0}$, is sufficiently far from the mass of 
the $\eta_b$~\cite{BABAR_etab}. 

We select events with exactly two oppositely charged tracks and a single energetic photon with 
a center-of-mass energy $E^*_{\gamma} \leq 0.2$ GeV, while allowing additional photons
with the center-of-mass energies below 0.2 GeV. We assign a muon mass hypothesis to the two tracks and require
that at least one is positively identified as a muon. We then perform a kinematic fit to the $\Upsilon$ 
candidate from the two muon candidates and the energetic photon. 

The backgrounds are dominated by the QED processes: the continuum background $e^+e^- \to \gamma \mu^+ \mu^-$
and the initial-state- radiation (ISR) production of $\rho^0$, $\phi$, $J/\psi$, $\psi(2S)$, and $\Upsilon(1S)$
vector mesons. We suppress the background contributions from the decays of $\rho^0 \to \pi^+ \pi^-$ final state 
in which the pion is misidentified as a muon by requiring both charged tracks are positively identified 
as muons in the range $0.5 \leq  m_{A^0} < 1.05$ GeV. We also suppress the decays of 
$\Upsilon(2S) \to \gamma_2 \chi_b(1P)$, $\chi_b(1P) \to \gamma_1 \Upsilon(1S)$ 
[$\Upsilon(3S) \to \gamma_2 \chi_b(2P)$, $\chi_b(2P) \to \gamma_1 \Upsilon(1S)$], where $\gamma_2$ 
is the secondary photon by requiring that no secondary photon above a center-of mass energy of 0.1 GeV 
[0.008 GeV] is present in the event. 

\begin{figure}[!htb]
\begin{center}
\vspace*{-0.1cm}
\includegraphics[height=12.0cm]{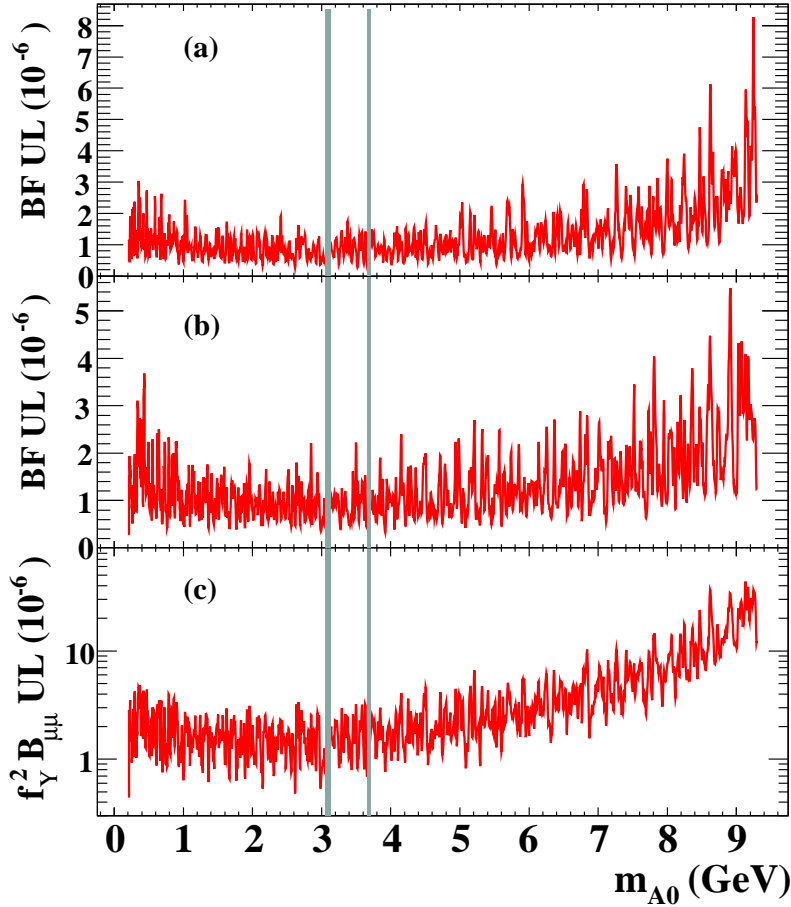}
\vspace*{-0.1cm}
\caption{The 90\% C.L. upper limits on the branching fractions of 
(a) ${\cal{B}}(\Upsilon(2S) \to \gamma A^0) \times {\cal{B}}_{\mu\mu}$, 
(b) ${\cal{B}}(\Upsilon(3S) \to \gamma A^0) \times {\cal{B}}_{\mu\mu}$, and (c) effective coupling 
$f^2_{\Upsilon} \times {\cal{B}}_{\mu\mu}$ as a function of $m_{A^0}$ (color online). 
The shaded ares show the regions around the $J/\psi$ and $\psi(2S)$ resonances excluded 
from the search.}
\label{fig:Egamma_nunu}
\end{center}
\end{figure}

Since we do not observe a significant excess of events in the range of $0.212 < m_{A^0} < 9.3$ GeV, 
we set upper limits on the branching fractions of ${\cal{B}}(\Upsilon(2S) \to \gamma A^0) \times 
{\cal{B}}_{\mu\mu}$ and ${\cal{B}}(\Upsilon(3S) \to \gamma A^0) \times {\cal{B}}_{\mu\mu}$, 
where ${\cal{B}}_{\mu\mu}$ is the branching fraction of $A^0 \to \mu\mu$.  
The limit vary from $0.26-8.3 \times 10^{-6}$ for ${\cal{B}}_{2S}$ and 
$0.27-5.5 \times 10^{-6}$ for ${\cal{B}}_{3S}$. 
The 90\% C.L. Bayesian upper limits are computed with a uniform prior and assuming a Gaussian
likelihood function. We set an upper limit on the coupling $f^2_{\Upsilon}(m_{A^0} = 
0.214$ GeV) $< 1.6 \times 10^{-6}$ at 90\% C.L., by assuming ${\cal{B}}_{\mu\mu} = 1$. This result is
significantly smaller than the value required to explain the HyperCP events as light Higgs 
production~\cite{HyperCP_2005, Light_higgs_2007}.  
Figure~\ref{fig:Egamma_nunu} shows the upper limits on the branching fractions
as a function of the mass $m_{A^0}$. 

We also measure the branching fractions of ${\cal{B}}(\Upsilon(2S) \to \gamma\eta_b) \times 
{\cal{B}}(\eta_b \to \mu^+\mu^-) = (-0.4 \pm 3.9 \pm 1.4) \times 10^{-6}$ and
${\cal{B}}(\Upsilon(3S) \to \gamma\eta_b) \times 
{\cal{B}}(\eta_b \to \mu^+\mu^-) = (-1.5 \pm 2.9 \pm 1.6) \times 10^{-6}$, where the first
uncertainty is statistical and the second is systematic, dominated by the the uncertainty
in $\Gamma(\eta_b)$. Using the \babar\ measurement of ${\cal{B}}(\Upsilon(2S) \to \gamma \eta_b)$ 
and ${\cal{B}}(\Upsilon(3S) \to \gamma \eta_b)$~\cite{BABAR_etab}, we derive 
${\cal{B}}(\eta_b \to \mu^+\mu^-) = (-0.25 \pm 0.51 \pm 0.33)\%$ and 
${\cal{B}}(\eta_b \to \mu^+\mu^-) < 0.9\%$ at 90\% C.L. This limit is consistent with the mesonic 
interpretation of the $\eta_b$ state~\cite{BABAR_nunu}.

\subsection{Search for \boldmath{$\Upsilon(3S) \to \gamma A^0$}, \boldmath{$A^0$} $\to$ \bf Invisible}

We search for the light scalar particle produced in single-photon decays of the $\Upsilon(3S)$ 
resonance through the process $\Upsilon(3S) \to \gamma A^0, A^0 \to$ invisible. We split the data sets 
into two broads energy ranges based on the energy of the highest-energy in the center-off mass 
in our electromagnetic calorimeter cluster. The high-energy region is $3.2 < E^*_{\gamma} < 5.5$ GeV,
where $E^*_{\gamma} = E_{com}/2$. The backgrounds in the high-energy region are dominated by 
the QED process such as $e^+e^- \to \gamma\gamma$, where the photon energy distribution for 
this process peaks. The low-energy region is $2.2 < E^*_{\gamma} < 3.7$. The backgrounds in 
this region are dominated by the radiative Bhabha events $e^+e^- \to \gamma e^+e^- \to$ in which
both electron and positron are misidentified.

We extract the yield of signal events using a likelihood fits to the distribution of the 
observable missing mass squared in the mass interval $ 0 < m_{A^0} < 6$ GeV:
\begin{eqnarray}
m^2_X = m^2_{\Upsilon(3S)} - 2 E^*_{\gamma} m_{\Upsilon(3S)},
\end{eqnarray}
where $m_X$ is the missing mass squared, $m_{\Upsilon(3S)}$ is the $\Upsilon(3S)$ mass, and
$E^*_{\gamma}$ is the photon energy in the $\Upsilon$ center-off-mass system. The $E^*_{\gamma}$ is given
by
\begin{eqnarray}
E^*_{\gamma}= \frac{m^2_{\Upsilon} - m^2_{A^0}}{2m_{\gamma}}
\end{eqnarray}
The current precise limit on the measurement of the branching fraction of ${\cal{B}}(\Upsilon \to \gamma X)$
with $X \to$ invisible is given by CLEO collaboration on $\Upsilon(1S)$~\cite{CLEO_1S}.  

In the high-energy range, the selection efficiency for signal is $10-11\%$ depending on the mass $m_{A^0}$,
and it is below $10^{-5}$ for $e^+e^- \to \gamma \gamma$ events, whereas in the low-energy range the
selection efficiency for signal is about 20\%. The signal PDF is described by the Crystal Ball function~
\cite{CB_function} that centered around the expected value of $m^2_X = m^2_m{A^0}$. We compare
the distributions of the simulated and reconstructed $e^+e^- \to \gamma \gamma$ events to determine
the the uncertainty of the PDF parameters.

\begin{figure}[!htb]
\begin{center}
\vspace*{-0.1cm}
\hspace*{-1.5cm}
\includegraphics[height=6.7cm]{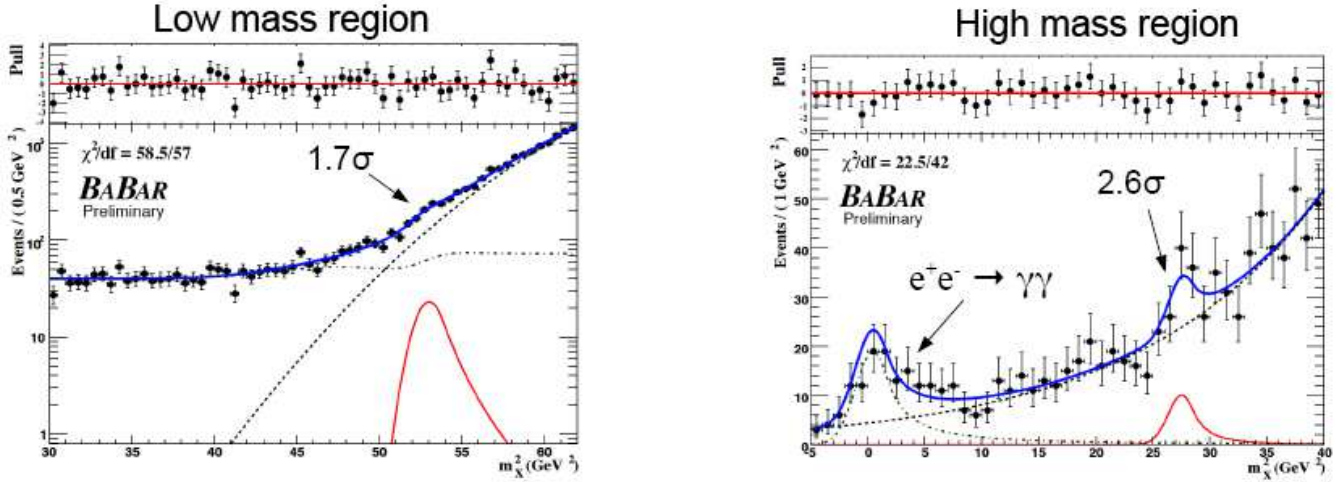}
\vspace*{-0.1cm}
\caption{[Left] Sample fit to the low-energy dataset ($83 \times 10^6~\Upsilon(3S)$) decays. The bottom plot shows the
data (solid points) overlaid by the full PDF curve (solid blue line), signal contribution with $m_{A^0} = 7.275$ GeV
(solid red line). [Right] Sample fit the high-energy dataset ($122 \times 10^6~\Upsilon(3S)$) decays. 
The bottom plot shows the data (solid points) overlaid by the full PDF curve (solid blue line), signal 
contribution with $m_{A^0} = 5.2$ GeV, $e^+e^- \to \gamma \gamma$ (dot-dashed green line), 
and the continuum background (black dashed line).}
\label{fig:Y3S_invisible}
\end{center}
\end{figure}

We determine the PDF as a function of mass $m_{A^0}$. In the low-statistics simulated samples of signal events, 
we exclude $6 < m_{A^0} < 7.8$ GeV and in the high-statistics simulated samples of signal events, 
we excluding the values of $m_{A^0} < 6$ GeV due to low statistics. The fit results give us signal yields of
$N_{sig} = 119 \pm 71$ ($1.7\sigma$) and $N_{sig} = 37 \pm 15$ ($2.6\sigma$) for the low-energy dataset 
and the high-energy dataset, respectively. The fit results on the low-energy range ($m_{A^0} = 7.275$ GeV) 
and the high-energy ($m_{A^0} = 5.2$ GeV) are shown in Fig.~\ref{fig:Y3S_invisible}.

We do not observe a significant 
excess of events above the background in the mass range of $0 < m_{A^0} < 7.8$ GeV., and we set 
90\% C.L. upper limit on the branching fraction of ${\cal{B}}(\Upsilon(3S) \to \gamma A^0) \times 
{\cal{B}}(A^0 \to$ invisible) at $(0.7 -31) \times 10^{-6}$ in the mass range of 
$m_{A^0} \leq 7.8$ GeV~\cite{Upsilon_3S_invisible}. These results are preliminary.

\subsection{Search for \boldmath{$\Upsilon(1S) \to$} \bf Invisible}
 
The nature of dark matter is one of the most challenging issues in the elementary particle physics.
Observation of the SM particles coupling to undetectable (invisible) particles might provide
information on dark matter candidates. The dark matter couples to the SM through the mediating boson.
This particle is expected to be a vector, scalar, or pseudoscalar. The dark matter to show up
dominantly in interactions with heavy fermions such as $b$ quarks~\cite{McElrath, Dermisek}.  

We select events in the invisible subsample by requiring exactly two tracks originating from 
the interaction point (IP) with opposite charge. These track have a center-off-mass momentum of 
$p^* < 0.8$ GeV/c and they are consistent with pions from dipion transition with the invariant mass
of $0.25 < M_{\pi\pi} < 0.95$ GeV/$c^2$. The dipion recoil mass is given by
\begin{eqnarray}
M^2_{rec} = s + M^2_{\pi\pi} - 2\sqrt(s)E^*_{\pi\pi},
\end{eqnarray}
where $M_{rec}$ is the reconstructed dipion recoil mass, $E^*_{\pi\pi}$ is the center-off-mass energy 
of the dipion system, and $\sqrt(s) = 10.3552$ GeV/$c^2$. The overall efficiency of the selection 
events is about 64\%.

\begin{figure}[!htb]
\begin{center}
\vspace*{-0.1cm}
\includegraphics[height=11.0cm]{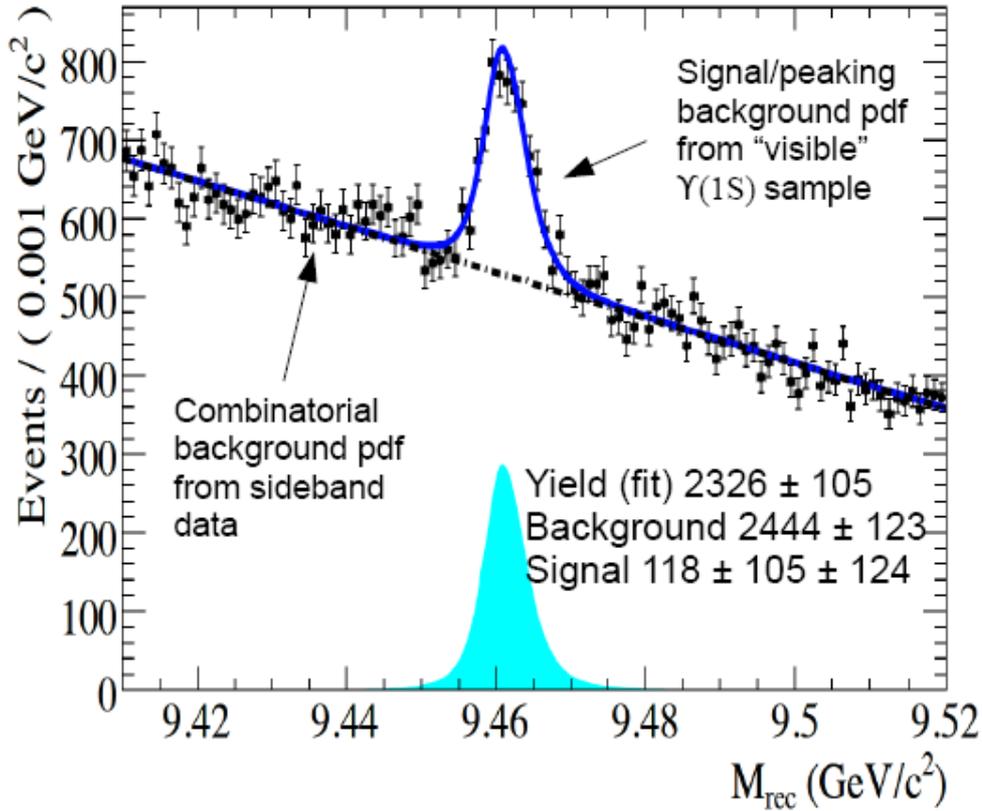}
\vspace*{-0.1cm}
\caption{The maximum likelihood fit to the dipion recoil mass for data in the invisible subsample.
Dash-dotted line is the nonpeaking background and the solid fitted-line is the peaking background.
The solid line is the total fit function.}
\label{fig:Upsilon1S_invisible}
\end{center}
\end{figure}

We collect the invisible subsample by selecting tree-track and four-track events using 
the dipion system. We search high-momentum tracks from $\Upsilon(1S)$ 
decays by requiring only one or two additional tracks originating from
the IP point, each with momentum  $p^* > 2.0$ GeV/$c^2$.  We then use our lepton identification to 
treat these tracks. If either both tracks passing electron-identification criteria the both tracks 
are treated as electron candidates; otherwise, both are treated as muon candidates.
Figure~\ref{fig:Upsilon1S_invisible} shows the $M_{rec}$ distribution from our invisible subsample.

We determine the parameters of the PDF by fitting the $M_{rec}$ distribution in the four-track
data subsample. We extract the peaking contribution by using an extended unbinned maximum 
likelihood fit and the non-peaking contribution using a first-order polynomial function.
The peaking background is estimated using Monte Carlo subsamples of 
1019 $\Upsilon(1S) \to e^+e^-$ events, 1007 $\Upsilon(1S) \to \mu^+\mu^-$ events, 
92 $\Upsilon(1S) \to \tau^+\tau^-$ events, and $2.9 \pm 1.3$ $\Upsilon(1S) \to$ hadrons events.

We search for evidence of the decays of $\Upsilon(1S)$ into undetectable final states recoiling
against the dipion system in the $\Upsilon(3S) \to \pi^+ \pi^- \Upsilon(1S)$, using a sample of 
$9.14 \times 10^6$ $\Upsilon(3S)$ mesons. We find no evidence for the decays of $\Upsilon(1S) \to$
invisible. We set an upper limit on the branching fraction of ${\cal{B}}(\Upsilon(1S) \to$ invisible)
$< 3.0 \times 10^{-4}$ at the 90\% C.L.~\cite{BABAR_1S_invisible}. 
  
\section{Summary}

We search for a new light non-Standard Model $CP$-odd Higgs boson in the 
$\Upsilon(3S)$, $\Upsilon(2S)$, and $\Upsilon(1S)$ systems. We also search for invisible
decays of the $\Upsilon(1S)$ system.  \babar\ recent measurements are important input 
to rule out much of the parameter space allowed by the light non-Standard Model $CP$-odd 
Higgs and other models. These measurements also provide new and stringent constraints on 
the light $CP$-odd Higgs boson. In addition the measurement of the branching fraction 
of $\Upsilon(1S) \to$ invisible is extremely important as a sensitive probe of dark matter.

\section{Acknowledgments}

The author would like to thank the organizers of QUARKS 2010, $16^{th}$ International Seminar 
on High Energy Physics, for their excellent program and kind hospitality. The supports from
the \babar\ Collaboration, the University of South Alabama, and the University of Mississippi 
are gratefully acknowledged. This work was supported by the U.S. Department of Energy under 
grant No. DE-FG02-96ER-40970.


\begin{thebibliography}{99}

\bibitem{LEP_2003}
R.~Barate {\it et al.} (LEP Working Group), Phys. Lett. B {\bf 565}, 61 (2003).

\bibitem{CDF_D0_2010}
T.~Aaltonen {\it et al.} (CDF and D0 Collaborations), Phys. Rev. Lett. {\bf 104}, 061802 (2010).

\bibitem{CMS_2010} 
V.~Khachatryan {\it et al.} (CMS Collaboration), arXiv:1005.3299 (2010);\\ 
V.~Khachatryan {\it et al.} (CMS Collaboration), JHEP 1002:041 (2010).

\bibitem{ATLAS_2010}
G.~Aad {\it et al.} (ATLAS Collaboration), CERN-PH-EP/2010-004, Phys. Lett. B 688 (2010).

\bibitem{Gunion_2006}
R.~Dermisek and J.~F.~Gunion, Phys. Rev. D {\bf 73}, 111701(R) (2010).

\bibitem{Babar_nim2002}
B.~Aubert {\it et al.} (\babar\ Collaboration), Nucl. Instrum. Methods Phys. Res., Sect. 
A {\bf 476}, 1 (2002).

\bibitem{BABAR_tautau}
B.~Aubert {\it et al.} (\babar\ Collaboration), Phys. Rev. Lett. {\bf 103}, 181801 (2009).

\bibitem{CLEO_tautau}
W.~Love {\it et al.} (CLEO Collaboration), Phys. Rev. Lett. {\bf 101}, 151802 (2008).

\bibitem{D0_tautau}
V.~M.~Abazov {\it et al.} (D0 Collaboration), Phys. Rev. Lett. {\bf 103}, 061801 (2009).

\bibitem{CB_function}
M.~J.~Oreglia, Ph.D. Thesis, SLAC-236 (1980);\\
J.~E.~Geiser, Ph.D. Thesis, Stanford University, SLAC-R-255 (1982);\\
T. Skwarnicki, Ph.D. Thesis, DESY F31-86-02 (1986).

\bibitem{PDG_2008}
C.~Amsler {\it et al.} (PDG Collaboration), Phys. Lett. B {\bf 667}, 1 (2008). 

\bibitem{BABAR_etab}
B.~Aubert {\it et al.} (\babar\ Collaboration), Phys. Rev. Lett. {\bf 101}, 071901 (2008).

\bibitem{HyperCP_2005}
H.~Park {\it et al.} (HyperCP Collaboration), Phys. Rev. Lett. {\bf 94}, 021801 (2005).

\bibitem{Light_higgs_2007}
X.~G.~He, J.~Tandean, and G.~Valencia, Phys. Rev. Lett. {\bf 98}, 081802 (2007).

\bibitem{BABAR_nunu}
B.~Aubert {\it et al.} (\babar\ Collaboration), Phys. Rev. Lett. {\bf 103}, 081803 (2009).

\bibitem{CLEO_1S}
R.~Ballest {\it et al.} (CLEO Collaboration), Phys. Rev. D {\bf 51}, 2053 (1995).

\bibitem{Upsilon_3S_invisible}
B.~Aubert {\it et al.} (\babar\ Collaboration), hep-ex/arXiv:0808.0017 (2008).

\bibitem{McElrath}
B.~McElrath, Phys. Rev. D {\bf 72}, 103508 (2005).

\bibitem{Dermisek}
R.~Dermisek, J.~F.~Gunion, and B.~McElrath, Phys. Rev. D {\bf 76}, 051105(R) (2007).  

\bibitem{BABAR_1S_invisible}
B.~Aubert {\it et al.} (\babar\ Collaboration), Phys. Rev. Lett. {\bf 103}, 251801 (2009).
%

\end{thebibliography}
\end{document}